\documentclass[%
 reprint,
superscriptaddress,
 amsmath,amssymb,
 aps,
floatfix,
]{revtex4-1}

\usepackage{graphicx}
\usepackage{dcolumn}
\usepackage{bm}
\usepackage{units}
\usepackage{braket}
\usepackage{epstopdf}
\usepackage{color}
\usepackage{float}
\usepackage{csquotes}

\usepackage[colorlinks, linkcolor=blue, citecolor=blue, urlcolor=blue, breaklinks=true]{hyperref}

\begin{document}

\title{Strategies for achieving high key rates in satellite-based QKD}

\author{Sebastian Ecker}
\email{sebastian.ecker@oeaw.ac.at}
\affiliation{Institute for Quantum Optics and Quantum Information (IQOQI), Austrian Academy of Sciences, Boltzmanngasse 3, 1090 Vienna, Austria.}
\affiliation{Vienna Center for Quantum Science and Technology (VCQ), Faculty of Physics, University of Vienna, Boltzmanngasse 5, A-1090 Vienna, Austria}

\author{Bo Liu}
\affiliation{Institute for Quantum Optics and Quantum Information (IQOQI), Austrian Academy of Sciences, Boltzmanngasse 3, 1090 Vienna, Austria.}
\affiliation{Vienna Center for Quantum Science and Technology (VCQ), Faculty of Physics, University of Vienna, Boltzmanngasse 5, A-1090 Vienna, Austria}
\affiliation{College of Advanced Interdisciplinary Studies, NUDT, Changsha, 410073, China}

\author{Johannes Handsteiner}
\affiliation{Institute for Quantum Optics and Quantum Information (IQOQI), Austrian Academy of Sciences, Boltzmanngasse 3, 1090 Vienna, Austria.}
\affiliation{Vienna Center for Quantum Science and Technology (VCQ), Faculty of Physics, University of Vienna, Boltzmanngasse 5, A-1090 Vienna, Austria}

\author{Matthias Fink}
\affiliation{Institute for Quantum Optics and Quantum Information (IQOQI), Austrian Academy of Sciences, Boltzmanngasse 3, 1090 Vienna, Austria.}
\affiliation{Vienna Center for Quantum Science and Technology (VCQ), Faculty of Physics, University of Vienna, Boltzmanngasse 5, A-1090 Vienna, Austria}

\author{Dominik Rauch}
\affiliation{Institute for Quantum Optics and Quantum Information (IQOQI), Austrian Academy of Sciences, Boltzmanngasse 3, 1090 Vienna, Austria.}
\affiliation{Vienna Center for Quantum Science and Technology (VCQ), Faculty of Physics, University of Vienna, Boltzmanngasse 5, A-1090 Vienna, Austria}

\author{Fabian Steinlechner}
\affiliation{Fraunhofer Institute for Applied Optics and Precision Engineering IOF, Albert-Einstein-Strasse 7, 07745 Jena, Germany}
\affiliation{Abbe Center of Photonics, Friedrich-Schiller-University Jena, Albert-Einstein-Str. 6, 07745 Jena, Germany}

\author{Thomas Scheidl}
\affiliation{Institute for Quantum Optics and Quantum Information (IQOQI), Austrian Academy of Sciences, Boltzmanngasse 3, 1090 Vienna, Austria.}
\affiliation{Vienna Center for Quantum Science and Technology (VCQ), Faculty of Physics, University of Vienna, Boltzmanngasse 5, A-1090 Vienna, Austria}

\author{Anton Zeilinger}
\affiliation{Institute for Quantum Optics and Quantum Information (IQOQI), Austrian Academy of Sciences, Boltzmanngasse 3, 1090 Vienna, Austria.}
\affiliation{Vienna Center for Quantum Science and Technology (VCQ), Faculty of Physics, University of Vienna, Boltzmanngasse 5, A-1090 Vienna, Austria}

\author{Rupert Ursin}
\email{rupert.ursin@oeaw.ac.at}
\affiliation{Institute for Quantum Optics and Quantum Information (IQOQI), Austrian Academy of Sciences, Boltzmanngasse 3, 1090 Vienna, Austria.}
\affiliation{Vienna Center for Quantum Science and Technology (VCQ), Faculty of Physics, University of Vienna, Boltzmanngasse 5, A-1090 Vienna, Austria}

\begin{abstract}
Quantum key distribution (QKD) is a pioneering quantum technology on the brink of widespread deployment. Nevertheless, the distribution of secret keys beyond a few 100 kilometers at practical rates remains a major challenge. One approach to circumvent lossy terrestrial transmission of entangled photon pairs is the deployment of optical satellite links. Optimizing these non-static quantum links to yield the highest possible key rate is essential for their successful operation. We therefore developed a high-brightness polarization-entangled photon pair source and a receiver module with a fast steering mirror capable of satellite tracking. We employed this state-of-the-art hardware to distribute photons over a terrestrial free-space link with a distance of 143 km, and extracted  secure key rates up to 300 bits per second. Contrary to fiber-based links, the channel loss in satellite downlinks is time-varying and the link time is limited to a few minutes. We therefore propose a model-based optimization of link parameters based on current channel and receiver conditions. This model and our field test will prove helpful in the design and operation of future satellite missions and advance the distribution of secret keys at high rates on a global scale.

\end{abstract}

\pacs{Valid PACS appear here}
\maketitle

Secure communications and data protection are the quintessential resources in an information-based society, with a wide range of applications such as financial transactions, ensuring personal privacy, and maintaining the integrity of critical infrastructure in the Internet of things. Quantum key distribution (QKD) allows to generate symmetric keys between distant parties, with a level of privacy that can be lower bounded from the very laws of physics. Flying qubits encoded in photons can be distributed up to distances of a few hundred kilometers in fibers ~\cite{Wengerowsky2020,Korzh2015,Inagaki2013}, while longer distances can be achieved by employing quantum repeaters ~\cite{Briegel1998,Sangouard2011}. Alternatively, optical satellite links have been proposed to overcome the distance limitations of ground-based transmission of photons \cite{Scheidl:2013,Ursin2009space}. The installation of quantum hardware on space platforms would also provide a platform for fundamental physics experiments ~\cite{Joshi:2018,Rideout:2012} and  radically new technologies such as quantum clock synchronization ~\cite{Giovannetti:2001, Ho:2009,Giovannetti:2001a} and quantum metrology~\cite{Ahmadi:2015}. While this is a technologically immensely challenging task, a number of experimental \cite{Ursin:2007,Fedrizzi:2009, Villoresi:2008, Pugh2017, Wang2013, Nauerth2013,Yin2012,Cao2013} and theoretical \cite{Scheidl:2013,Bonato2009} studies have established the feasibility of this vision with state-of-the-art technology available on ground and certified for operation in space. Consequently, in what has been called the \textit{quantum space race} \cite{Jennewein2013}, a number of international research groups in Canada, China, Europe, Japan and Singapore are pursuing first missions involving space links ~\cite{Khan2018,Bedington2017}, with first dedicated satellite transmitter payloads successfully launched into space~\cite{Yin:2017,Liao:2017,Takenaka:2017,Liao2017space,yin2020entanglement}.

\begin{figure}[H]
	\centering
	\includegraphics[width=\columnwidth]{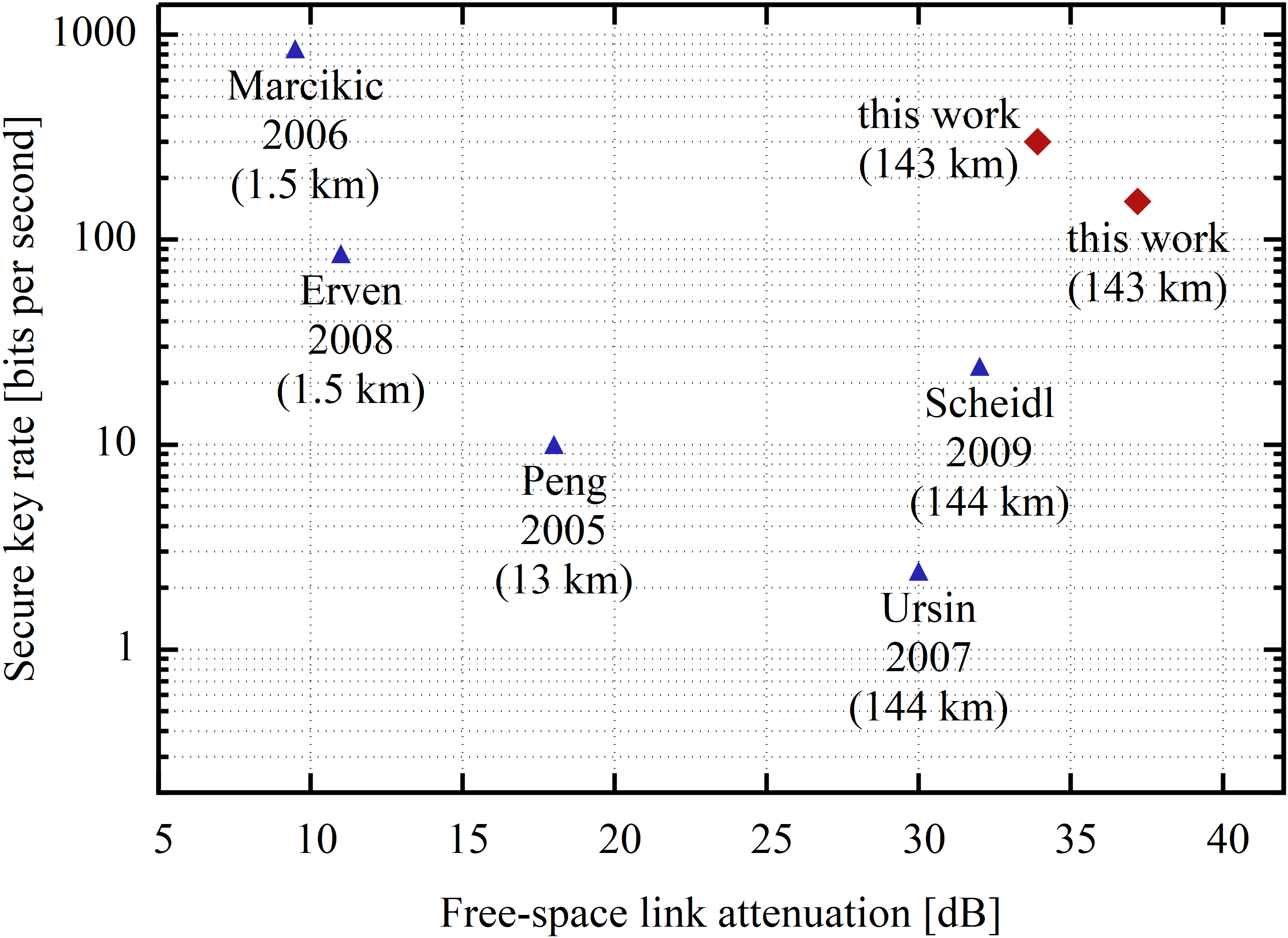}
	\caption{\textbf{Comparison of secure key rates}. Only published QKD field trials over a free-space link employing polarization-entangled photon pairs are compared. The horizontal axis corresponds to the attenuation caused by the free-space link and the receiver optics. Each blue data point represents a published experiment: Peng 2005 \cite{Peng:2005}, Marcikic 2006 \cite{Marcikic:2006}, Ursin 2007 \cite{Ursin:2007},  Erven 2008 \cite{Erven:2008}, Scheidl 2009 \cite{Scheidl:2009}; the red data points represent the two highest key rates achieved in this work. Apart from \cite{Erven:2008} and \cite{Peng:2005}, all experiments employ a single free-space link.}
	\label{fig:comparison}
	
\end{figure}

\begin{figure*}
	\centering
	\includegraphics[width=1\textwidth]{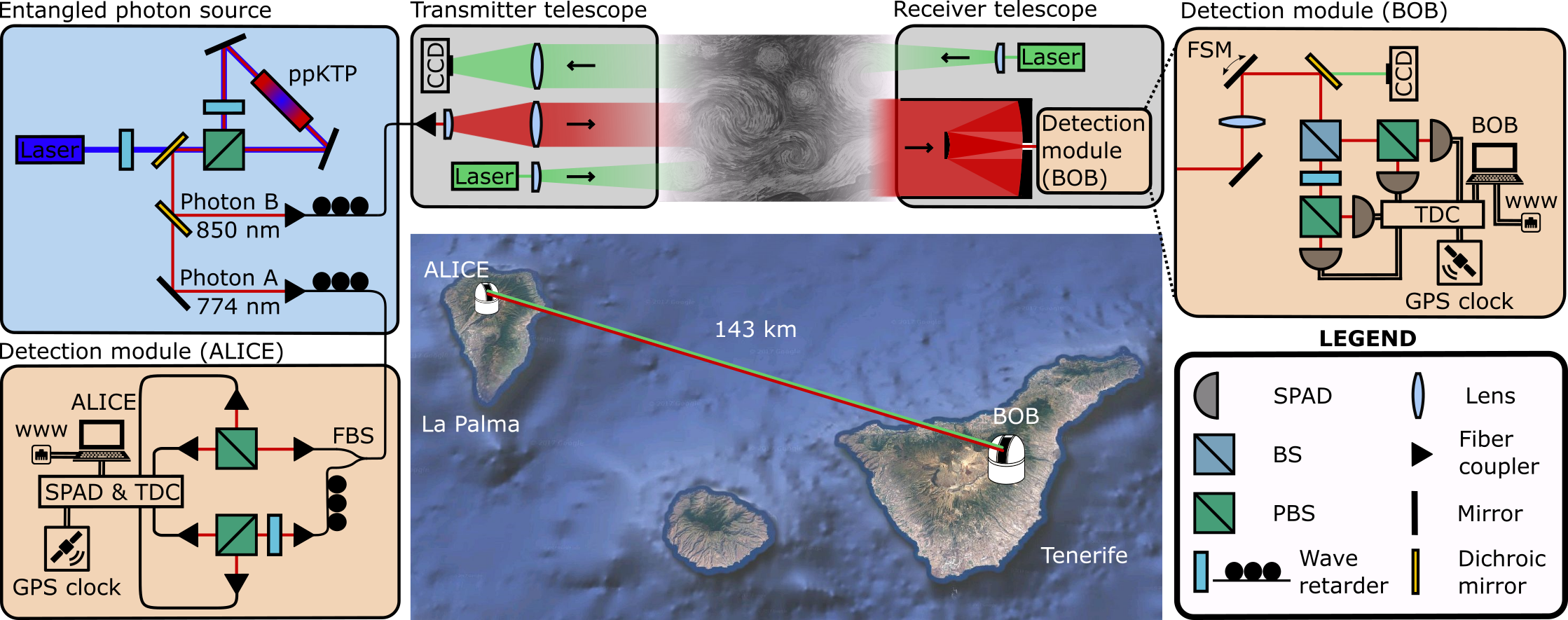}
	\caption{\textbf{Sketch of the experimental setup.} Alice and Bob were located on the Canary islands of La Palma and Tenerife, respectively, amounting to a free-space link distance of $\unit[143]{km}$. Polarization-entangled photon-pairs were produced in SPDC by bidirectionally pumping a ppKTP crystal placed in the center of a Sagnac interferometer. Both Alice's and Bob's measurements were preceded by a random polarization basis choice realized by non-polarizing beamsplitters (BS). The measurement in one of two mutually unbiased polarization bases was achieved by polarizing beamsplitters (PBS), followed by detection with single-photon avalanche diodes (SPAD). All detection events were recorded with time-to-digital converters (TDC) disciplined by pulse-per-second signals from GPS clocks. Storage and post-processing of the timestamps was done with PCs, which were both connected to the Internet, constituting the classical communication channel. For the purpose of pointing, acquisition and tracking, the quantum channel was bidirectionally overlapped with two green \unit[532]{nm} beacon lasers. Map data \copyright 2017 Google.}
	\label{fig:setup}
\end{figure*}

The quantum space race has also served as a driver for the development of robust quantum technology~\cite{Tang2016,Tang2016b,Steinlechner:2016}, with the recent years seeing tremendous advancements in quantum nonlinear optics, entangled photon generation methodology and single-photon detection. In light of these technological advances, a re-evaluation of the improved performance characteristics of QKD via representative free-space links is critical, especially concerning the increase in secure key rate compared to earlier field trials \cite{Marcikic:2006, Erven:2008, Peng:2005,Scheidl:2009,Ursin:2007}, illustrated in Fig.\ \ref{fig:comparison}. Field tests based on prepare-and-measure schemes have not been included in this comparison, though we would like to mention that both terrestrial \cite{Schmitt-Manderbach2007} and satellite-based  \cite{Liao:2017,Liao2018} studies have demonstrated decoy-state key exchange over free-space links at high rates. Entanglement-based QKD protocols remove the need to trust the source on the satellite in a dual downlink scenario.

In this article we report on a state-of-the-art experimental feasibility study for entanglement-based satellite QKD between the islands of La Palma and Tenerife. In order to be forearmed for future satellite down-link experiments we have developed a quantum ground receiver for polarization-based QKD protocols, which is compatible with most existing optical ground stations with satellite tracking capabilities. We distributed entangled photons from an ultra-bright source over a 143-km-long atmospheric free-space link with $>\unit[40]{dB}$ total channel loss, which is comparable to the average down-link loss for a low earth orbit (LEO) satellite pass \cite{Yin2017}. Additionally, we measured the background sky noise for a typical pass of a LEO satellite. Under these conditions we implemented the BBM92 protocol \cite{Bennett1992}, supplemented by error correction and privacy amplification, which yielded secure key rates up to 300 bits per second (bps) including finite-size-effects \cite{Fung2010}, which ranks amongst the highest key rates over a free-space channel with $>\unit[40]{dB}$ total channel loss. This was achieved by adapting the pair production rate to the current channel attenuation. The same model-based optimization which has proven successful for our terrestrial free-space link was subsequently used to estimate achievable secure key rates in a LEO dual downlink scenario \cite{Yin:2017}. 
Our results provide updated estimates for entanglement-based satellite-ground QKD and provide a valuable guideline for the design of future space missions.

\section*{Results}
\noindent
\textbf{Setup of Field Trial.}
The goal of QKD is the distribution of a secure key between Alice and Bob, which is subsequently used for the symmetric encryption of private messages. In our field trial, these communicating parties  were located on the Canary islands of La Palma (Alice) and Tenerife (Bob). A source of polarization-entangled photon pairs (photon A and B) was placed near the Jacobus Kapteyn Telescope of the Roque de los Muchachos Observatory on La Palma at an altitude of \unit[2360]{m}. Photon A was sent to Alice's detection module via a few meters of single-mode fiber, while photon B was transmitted through a \unit[143]{km}-long free-space link, after which it was collected by a reflector telescope in the Optical Ground Station (OGS) of the Observatorio del Teide at an altitude of \unit[2400]{m} (see Fig.\ \ref{fig:setup}). In order to compensate for slow beam drifts caused by the atmosphere, the transmitter and receiver telescopes were actively guided towards each other based on bidirectionally overlapped green beacon lasers. Both Alice's and Bob's detection modules are comprised of a random polarization basis choice, realized by a beam splitter, followed by a projection on one of two conjugate polarization bases. The different parts of the experimental setup are described in more detail in the `Methods' section. In the following, we will elaborate on the performance of the free-space link and on the distillation of a secure key from priorly distributed entangled photon pairs.
\begin{figure}[H]
	\centering
	\includegraphics[width=0.48\textwidth]{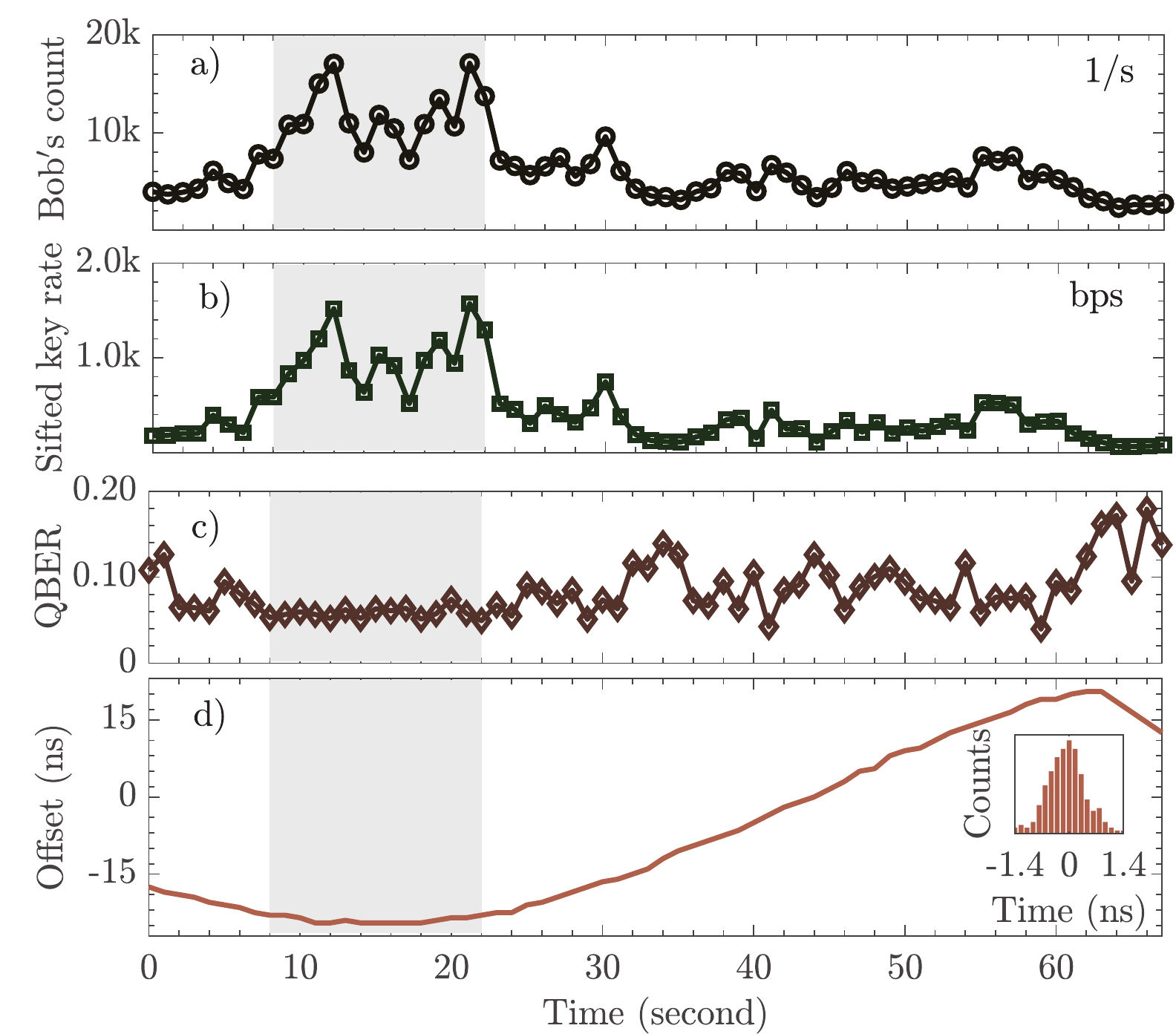}
	\caption{\textbf{Performance of the free-space link over time.} One of the photons is detected in Alice's laboratory at a rate of $\unit[13.3]{Mcps}$, while the polarization-entangled partner photon is sent over the free-space link. a) Due to atmospheric turbulence, the rate of single photons Bob receives is varying over time. Each data point corresponds to an integration time of $\unit[1]{s}$. b) Sifted key rate and c) Quantum bit error rate (QBER) of the BBM92 protocol \cite{Bennett1992}. d) Relative clock drift between Alice and Bob around the time-of-flight offset of $\unit[478.12]{\mu s}$. The inset depicts the normalized correlation peak of two-photon detections after time synchronization with a resolution of \unit[156]{ps}.}
	\label{fig:SKR_Experiment}
\end{figure}
\noindent
\textbf{QKD over the free-space link.}
A well-characterized communication channel is pivotal for long-distance quantum communication. Different from optical links over glass fibers, free-space links experience time-varying loss due to atmospheric turbulence. This results in a single-photon count rate at Bob which is fluctuating in time (Fig. ~\hyperref[fig:SKR_Experiment]{\ref*{fig:SKR_Experiment}a)}), while Alice detects her partner photon with a constant single-photon rate of $\unit[13.3]{Mcps}$. After identifying the coincidences between these vastly different single-photon count rates, Alice and Bob share a raw key. In basis reconciliation, they classically communicate their random basis choice and discard all events with unmachted measurement bases, resulting in a sifted key (Fig.~\hyperref[fig:SKR_Experiment]{\ref*{fig:SKR_Experiment}b)}). From this sifted key, a quantum bit error rate (QBER) can be estimated (Fig.~\hyperref[fig:SKR_Experiment]{\ref*{fig:SKR_Experiment}c)}). The QBER is higher at times of low channel transmission, since the constant background count rate leads to a lower signal-to-noise ratio. QKD at high rates requires a precise time-synchronization between the communication parties. We harness the strong intensity correlations of the photon pairs in order to compensate for relative clock drifts (Fig.~\hyperref[fig:SKR_Experiment]{\ref*{fig:SKR_Experiment}d)}). The retrieved two-photon correlation peak (see inset) exhibits a width which corresponds to the timing-jitter of the detection and electronic systems. Photon pair detections are subsequently identified using a coincidence window of 1 ns. For these post-processing steps we use a custom software which is capable of coincidence retrieval on a sub-second timescale. This enables us to align free-space links based on coincidence detections. 
\begin{figure}[H]
	\centering
	\includegraphics[width=0.48\textwidth]{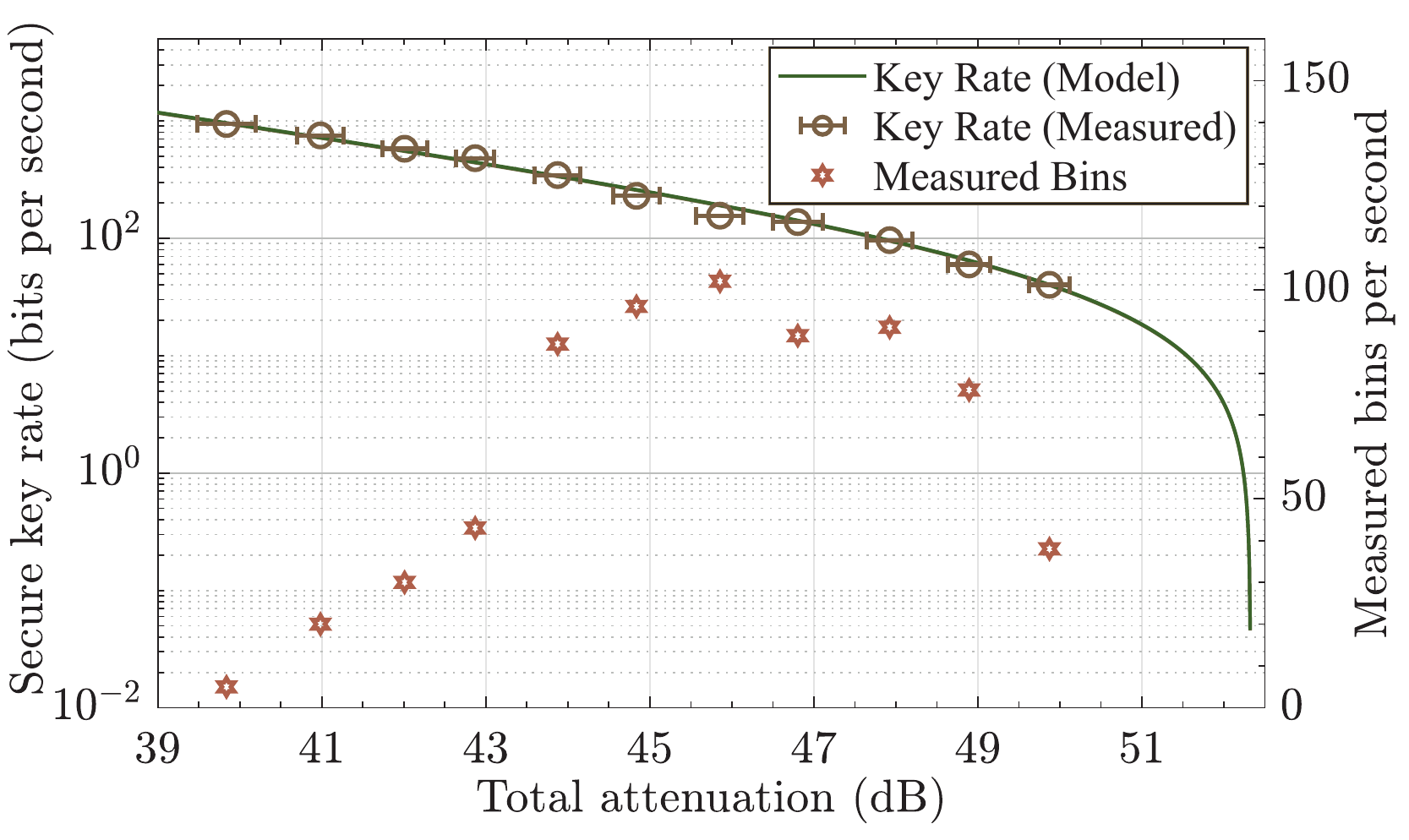}
	\caption{\textbf{Variation of the secure key rate as a result of atmospheric turbulence.} Beam wander and beam spreading across the free-space link lead to a fluctuating channel attenuation. This is quantified by binning a \unit[148]{s}-long measurement in \unit[100]{ms}-intervals, in each of which the loss is computed via Eq.~\eqref{eq:loss}. The starlike data points correspond to the frequency of these bins, while the circular data points correspond to the average secure key rate within 1-dB-loss-bins, where the horizontal error bars indicate the standard deviation of the corresponding average loss. The link is modelled according to Eq.~\eqref{equ:KeyRate} and the resulting rate is plotted (solid line).  See Supplementary Table 2 for further details.}
	\label{fig:SKR_channel_loss}
\end{figure}
In order to estimate the length of the secure key $N_\text{f}$, we use the bound 

\begin{equation}
\begin{split}N_\text{f}&\geq N_\text{s}^{\text{z}}\left[ 1 - H_2\left(E^{\text{ph},\text{z}}_{\tilde{\mu}}\right) - f\left(E_{\tilde{\mu}}^\text{z}\right)H_2\left(E_{\tilde{\mu}}^\text{z}\right)\right]\\
&+N_\text{s}^{\text{x}}\left[ 1 - H_2\left(E^{\text{ph},\text{x}}_{\tilde{\mu}}\right) - f\left(E_{\tilde{\mu}}^\text{x}\right)H_2\left(E_{\tilde{\mu}}^\text{x}\right)\right],
\label{equ:KeyRate}
\end{split}
\end{equation}

which is derived in~\cite{Takesue2009,Maentangle2007,Fung2010}.
Here,  $N_\text{s}^{\text{z}}$ ($N_\text{s}^{\text{x}}$) is the sifted key length in the $Z(X)$-basis. The QBER in the two bases are  $E_{\tilde{\mu}}^\text{z}=\unit[6.6]{\%}$ and $E_{\tilde{\mu}}^\text{x}=\unit[7.07]{\%}$, while $E^{\text{ph},\text{z}}_{\tilde{\mu}}$ ($E^{\text{ph},\text{z}}_{\tilde{\mu}}$) denotes the estimated phase error rate in the z(x)-basis. All error rates are evaluated as arguments of the binary entropy function $H_2(x)$, with an error correction efficiency for the Low-density parity-check (LDPC) code of $f(x)=1.2$. Allowing for a failure probability of $\varepsilon_{\text{ph}}=10^{-5}$ and disregarding finite-size effects, the average key rate is \unit[71.8]{bps} over the whole measurement time of $\unit[68]{s}$ (see Supplementary Table 1). Within a $\unit[15]{s}$-measurement interval (grey region in Fig.~\ref{fig:SKR_Experiment}), the average secure key rate is $\unit[300]{bps}$ with an average attenuation of \unit[38.72]{dB} (from the source to Bob's detectors). 

In a next step, we investigate the relationship between the secure key rate and the link loss. To this end, the measurement data is sliced in $\unit[100]{ms}$ time-bins and the attenuation of Bob's channel within each time-bin is obtained by

\begin{equation}
    \alpha_{\mathrm{ch}}[\text{dB}] = -10\log_{10}\left(\frac{R_{\text{CC}}-R_\text{A}R_\text{B}\tau_{\text{cw}}}{R_\text{A}}\right),
\label{eq:loss}
\end{equation}

where $R_{\text{CC}}$ is the coincidence count rate, $R_\text{A}$ and $R_\text{B}$ are the single count rates of Alice and Bob and $\tau_{\text{cw}}$ is the coincidence window. The total channel attenuation is given by adding another 4.8 dB for Alice's channel loss. After grouping the losses in 1-dB-intervals, we calculate the average loss and the corresponding secure key rate within this intervals as shown in Fig.\ \ref{fig:SKR_channel_loss}. The secure key rates obtained from the measurement data coincide with the model in Eq.~\eqref{equ:KeyRate} without any free parameters. 
\begin{figure}[t]
	\centering
	\includegraphics[width=0.48\textwidth]{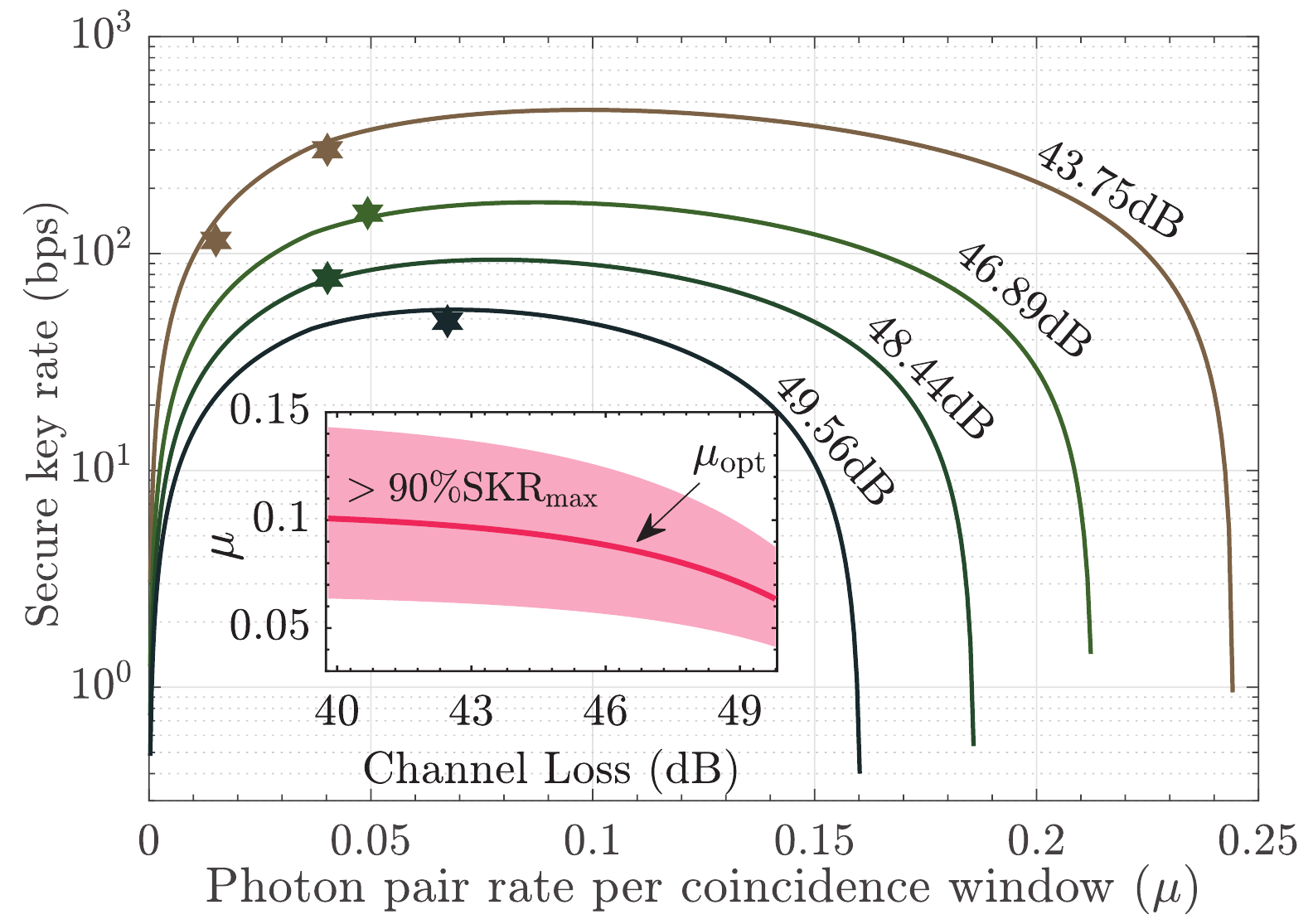}
	\caption{\textbf{Achievable secure key rates over a single free-space link.} The solid lines are model calculations for four different average  total link attenuations corresponding to five of our measurements (stars). While the maximal achievable secure key rate SKR\textsubscript{max} decreases for increasing attenuation, the optimal photon pair rate $\mu_\text{opt}$ to achieve SKR\textsubscript{max} shifts to lower pair rates. The inset illustrates this tendency and highlights the tolerance ($\geq\unit[90]{\%}$ of SKR\textsubscript{max}) to pair rate fluctuations around $\mu_\text{opt}$.} 
	\label{fig:SKR_Mu}
\end{figure}
\\\\
\noindent\textbf{Optimization of the key rate.} While the loss across the channel is not under our control, both the source and the receiver offer adjustable parameters, the most important of which are the emitted pair rate and the coincidence window. For our setup the maximal key rate is achieved with a coincidence window of 1 ns, which does not necessarily coincide with the minimal QBER. In general, the optimal coincidence window depends on the timing jitter of the detection and electronic systems and the clock synchronization accuracy. Adjusting the emitted photon pair rate $\mu$ is easily achieved by tuning the power of the pump laser (see `Methods' section for experimental details) . Interestingly, the secure key rate is not proportional to the pair rate of the source due to limitations in photon detection. These limitations include accidental coincidence counts and nonlinear detector responses such as afterpulsing \cite{Ziarkash2018} and dead-time effects, which we include in our model. As a consequence, higher pair rates result in higher QBERs, which lead to a decline in secure key rate after a loss-specific optimal pair rate $\mu_\text{opt}$ (see Fig.\ \ref{fig:SKR_Mu}) \cite{Holloway2013}. Operating the link at $\mu_\text{opt}$ leads to the highest achievable secure key rate SKR\textsubscript{max}. While the plateau around SKR\textsubscript{max} is very broad for low average losses, adjusting the source to $\mu_\text{opt}$ for high average link losses is more delicate (see inset). The highest key rate we could have achieved is close to 500 bps, which is $\sim 60\%$ higher than the key rate we observed. Unfortunately, most of our measurements weren't operated at $\mu_\text{opt}$ due to a lack of a detailed model at the time of the measurements. However, the measurements we performed (see Supplementary Table 3) coincide with our model predictions, again indicating the validity of our model. 

\begin{figure}[t]
	\centering
	\includegraphics[width=0.48\textwidth]{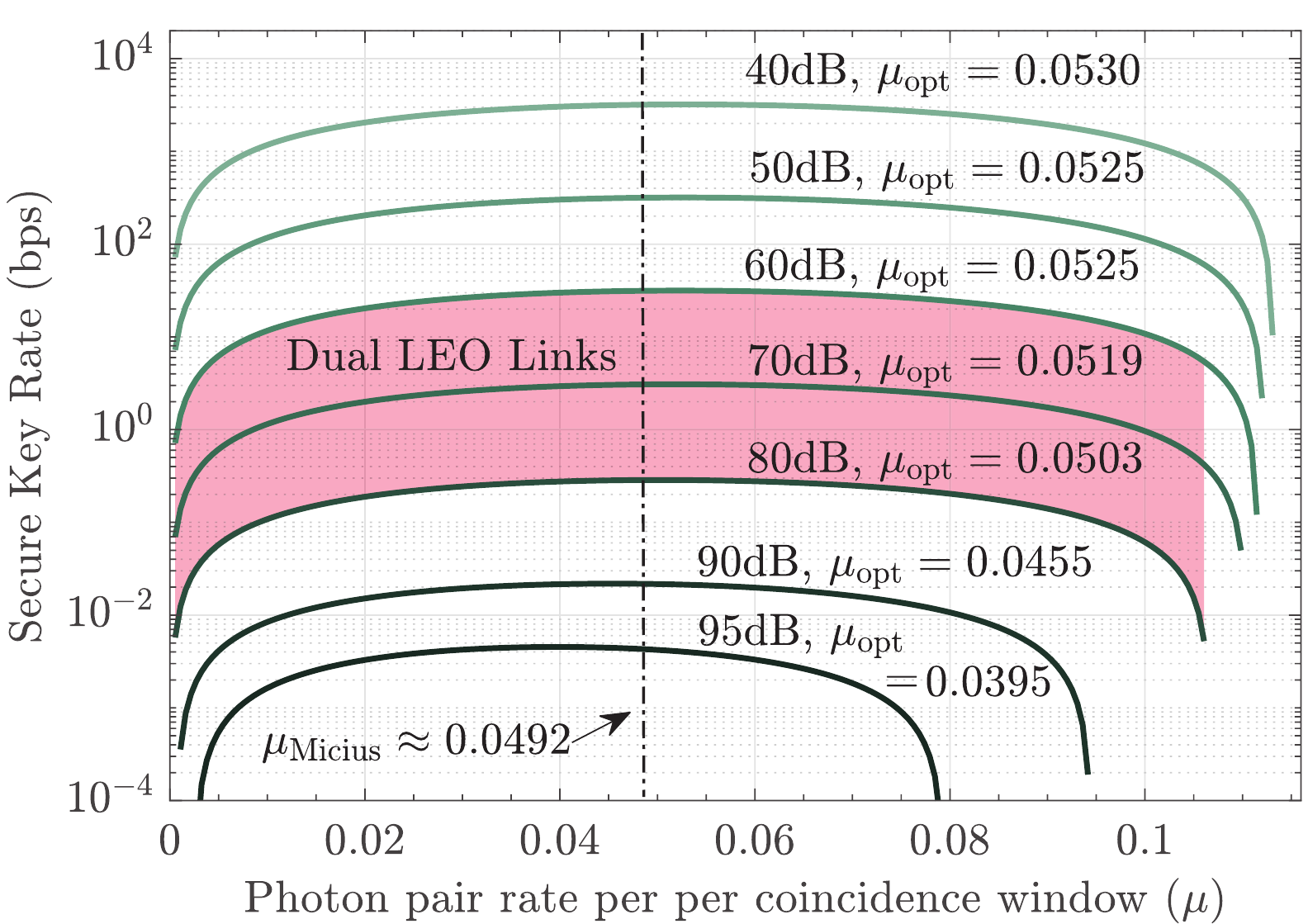}
	\caption{\textbf{Achievable secure key rates in a symmetric dual downlink scenario.} The model parameters are taken from the Micius dual downlink experiment \cite{Yin:2017}. Each channel attenuation has a specific optimal pair rate $\mu_\text{opt}$ which yields the maximal achievable secure key rate. Additionally, the photon pair rate $\mu_\text{Micius}$ emitted by Micius' entangled source is marked. The coloured region corresponds to the link attenuation range of the Micius dual downlink.}
	\label{fig:SKR_Mu_DualLink}
\end{figure}

With a valid model of our free-space channel at hand, we can now predict secure key rates achievable in a LEO dual downlink scenario. From the measurement data of the Micius dual downlink experiment \cite{Yin:2017,yin2020entanglement} we know which losses and background count rates to expect. Based on specifications of the Micius satellite and the Chinese ground stations, the dual downlink is modelled (see Supplementary Information) and yields the secure key rates plotted in Fig.~\ref{fig:SKR_Mu_DualLink}. The link loss is influenced by the elevation angle of the satellite and ranges from 60 dB to 80 dB. For two symmetrical high-loss channels, the optimal pair rate $\mu_\text{opt}$ is almost independent of the total channel attenuation and with a coincidence window of 2.5 ns and 5.9 million photon pairs per second, the Micius dual downlink is operated close to the optimal pair rate $\mu_\text{opt}$ (see Fig.~\ref{fig:SKR_Mu_DualLink}).

\section*{Discussion}

We presented a QKD field trial over a free-space channel with a link distance of $\unit[143]{km}$. This constitutes a worst-case scenario for satellite-based QKD with a LEO downlink in terms of channel attenuation \cite{Yin:2017}, astronomical seeing \cite{Ursin:2007} and sky noise. One of the main challenges of a satellite downlink, namely pointing and tracking in orbit, could not be simulated on our terrestrial free-space link. We used the polarization degree of freedom to distribute entangled photon pairs between the communicating parties and extracted a secure key with the BBM92 protocol. By employing state-of-the-art single-photon detectors and an ultra-bright source of entangled photon pairs, we obtained a secure key rate of \unit[154]{bps} over \unit[85]{s} with an average total channel loss of \unit[46.9]{dB} and a record secure key rate of \unit[300]{bps} over 15 s with an average total channel loss of \unit[43.5]{dB}, including finite-size-effects. Continuous key generation over longer timescales is irrelevant for a LEO scenario, since typical satellite passes are limited by a few minutes of link time and as opposed to our terrestrial link, the downlink-attenuation is not heavily fluctuating on a second timescale \cite{Yin:2017,Yin2017}. The obtained secure key rates agree with our models, which factor in all relevant source, detector and channel characteristics. Our results are also comparable to the predicted average key rate of $\sim\unit[350]{bps}$ in the case of higher triggering efficiencies in a Micius satellite downlink \cite{Yin2017}.

Although the attenuation in a satellite down-link is set by the orbit of the satellite, both the transmitter as well as the receiver can be optimized according to current link conditions. 
An important parameter in this regard is the photon pair rate emitted by the source, which can be optimized depending on the channel loss and the detector characteristics. The elevation angle of a LEO satellite leads to channel attenuations ranging from 29 dB to 45 dB \cite{Yin2017}, which makes an adjustment of the photon pair rate on the satellite indispensable in order to maximize the number of secure bits exchanged within a short LEO satellite pass. Monitoring this time-varying attenuation could be accomplished by measuring the power of the beacon or synchronisation laser at the satellite. This information can be used to adjust the pump power of the spaceborne photon pair source on the timescale of several ms in order to adapt to current atmospheric conditions. However, small deviations from the optimal pair rate do not lead to a significant change in secure key rate, which relaxes the requirements on the output power stabilisation of diode lasers launched into space substantially. 

Additionally to the experimental investigation of a single free-space link, we modelled the dual downlink from Micius, which yielded optimal photon pair rates close to the actual pair rates achieved by Micius. In the case of two nearly symmetrical high-loss channels, the optimal photon pair rate is not changing significantly as a function of the channel loss. However, since Micius and follow-up satellites will connect to an increasing number of ground stations around the globe, it is important to note that each receiver will have different frequency standards and detector characteristics, which yields different optimal coincidence windows and photon pair rates. As opposed to static fiber-based networks, a future satellite-based QKD network will therefore heavily rely on models dynamically predicting optimal parameters based on current link and receiver conditions.

In our study we intended to showcase obtainable secure key rates with state-of-the-art sources and receivers. A number of technological advancements must be accomplished in order to significantly increase these key rates.
While the brightness of photon-pair sources based on quasi-phase matching in SPDC is sufficient to operate both single and dual down-links at the maximal key rate, as shown in this study, this is only true for detection with semiconductor single-photon detectors. The advancements in superconducting nanowire single-photon detectors (SNSPDs) \cite{Natarajan2012} are promising for high-loss quantum communication, since they are characterised by short dead-times, negligible dark counts, very high detection efficiencies and low timing-jitter \cite{Dobrovolskiy2017}. However, these detectors can not be straightforwardly employed for free-space applications, since they currently only exist in single-mode coupling. In order to make SNSPDs free-space compatible, the multi-mode beam could be either coupled in single-mode fibers with adaptive optics or directly impinge on the nanowire through a vacuum viewport. Due to their better timing resolution and higher maximal count rates, these detectors will increase the optimal pair rate significantly. For typical SNSPD specifications (see Supplementary Table 4) the optimal pair rate in a dual downlink is \unit[814]{Mcps}. Recent developments on ultra-bright photon pair sources, however, show that such values are well within reach \cite{Steinlechner12,PhysRevLett.120.140405}.

In order to decrease the rate of detected background photons, tight filtering in the spatial and spectral domain of the photons is inevitable, which has already been demonstrated for terrestrial free-space links \cite{Peloso:2009,Liao2017daylight}. Further measures to mitigate noise include adaptive optics \cite{weyrauch2002fiber} and ultra-narrowband photon pair sources \cite{Tsujimoto2018}. Frequency multiplexing of polarization-entangled photon pairs \cite{Aktas2016} is an efficient way to increase the secure key rate over free-space links, since each frequency channel can be operated at the optimal pair rate, while the  added complexity is shifted to the receivers on ground. 
Another possible avenue for high-rate QKD is the distribution of high-dimensional entanglement over free-space links \cite{Steinlechner2017}, which increases the per-photon information capacity \cite{Ali-Khan2007,Zhong2015} and enhances the resilience against noise \cite{PhysRevX.9.041042}. We hope that this work will prove beneficial in the design of future LEO missions. The employment of state-of-the-art quantum sources and receivers together with the dynamical prediction of optimal parameters in orbit enable high key rates and pave the way towards QKD on a global scale.

\section*{Methods}
\noindent
\textbf{Ultra-bright Entangled Photon-pair Source.}
The source of photon pairs was based on type-0 spontaneous parametric down-conversion (SPDC). In order to achieve polarization entanglement, a periodically poled $\text{KTiOPO}_{4}$
(ppKTP) crystal was placed within a polarization Sagnac interferometer \cite{Fedrizzi:2007,Kim2006,Steinlechner14}. The $\unit[20]{mm}$-long ppKTP crystal was bi-directionally pumped with a continuous-wave diode laser tuned to $\unit[405]{nm}$, tightly focused on the crystal for increased brightness. By tuning the temperature of the crystal, photons with non-degenerate central wavelengths $\lambda_\text{A}=\unit[774]{nm}$ and $\lambda_\text{B}=\unit[840]{nm}$ were produced. Using dichroic filters, the photon pair was separated in two distinct spatial modes and coupled into single-mode fibers.

After longpass and $\unit[3]{nm}$-bandpass filtering, the source produced a local two-fold detection rate of $\unit[\sim 280]{kcps}$ per $\unit[]{mW}$ of pump power with a symmetric heralding efficiency of $\unit[\sim 33]{\%}$.
Polarization entanglement was verified by measuring the second order interference Visibilities V in the rectilinear (H/V) and diagonal (D/A) polarization basis. Typical experimental values yielded $V_{\text{H/V}}=\unit[99]{\%}$ and  $V_{\text{D/A}}=\unit[98.5]{\%}$ measured locally, which corresponds to a Fidelity F = $(V_{\text{H/V}}+V_{\text{D/A}})
/2\simeq 0.987$ with the maximally-entangled Bell state $\ket{\Phi^-}=1/\sqrt{2}\left(\ket{H_\text{A}V_\text{B}}-\ket{V_\text{A}H_\text{B}}\right)$. 
\\\\
\noindent
\textbf{Detection Modules.}
Both Alice's and Bob's measurements were preceded by a random polarization basis choice implemented by non-polarizing 50/50 beamsplitters (BS). In the output ports of the BS, polarizing beamsplitters (PBS) projected the photons in one of two mutually unbiased polarization bases (H/V- or z-basis and D/A- or x-basis). Single photon counting was accomplished by means of 4 single-photon avalanche diodes (SPAD) for each detection module. They were fiber-coupled in Alice's detection module (Excelitas Technologies SPCM-800-11), and free-space-coupled in Bob's detection module (ID Quantique ID120, diameter of active detector area = $\unit[500]{\mu m}$ ). The timestamps of the single-photon detection events were locally recorded with time-to-digital converters (AIT TTM8000) and stored on hard drives. Due to the relative drift between the local clocks, the  time tagging units were disciplined by pulse-per-second signals from GPS clocks, which were utilized for coincidence retrieval in post-processing. Key extraction from the raw timestamps was accomplished with PCs, which were both connected to the Internet, constituting the classical communication channel.
\\\\
\noindent
\textbf{Free-space channel.}
The free-space channel consisted of a transmitter telescope, which shaped the beam for long-distance transmission, followed by a $\unit[143]{km}$-long free-space link and a receiver telescope, which collected the photons for subsequent measurement. In order to minimize diffraction at the transmitter telescope aperture, the sending lens (diameter $= \unit[70]{mm}$, focal length $=\unit[280]{mm}$) was only partially illuminated by the photons emanating from the single-mode fiber.
In a diffraction-limited case this would lead to a spotsize at the receiver telescope of $\unit[1.3]{m}$ in diameter. However, beam wandering and beam spreading caused a widening of the beam  due to propagation through turbulent air. Apart from these geometrical losses, the photons experienced absorption and scattering across the free-space link. These effects summed up to a total average link loss from the fiber to the receiver telescope of $\unit[33]{dB}$. The receiver optics was estimated to add another $\unit[6]{dB}$ of optical loss. After the free-space link, the photons were collected by a $\unit[1]{m}$ Ritchey-Chr\'etien reflector telescope. Additionally to signal photons from Alice, Bob collected background photons with an average rate of \unit[450]{cps} per detector channel. We also measured the sky noise rate of a typical LEO satellite pass with the same quantum receiver (see Supplementary Figure 1), which turned out to be slightly lower than the terrestrial link background rate. 

For the purpose of pointing, acquisition and tracking, the free-space quantum channel was bidirectionally overlapped with green \unit[532]{nm} beacon lasers, which were imaged by charge-coupled devices (CCD) on both the transmitter and receiver side. The green light impinging on Alice's CCD was used for a closed-loop tracking system, compensating beam drifts on the timescale of several seconds caused by varying temperature and humidity gradients across the free-space link. 

Since the OGS and the mounted detection module was designed for satellite tracking, the CCD on Bob's side was in a closed loop with a fast steering mirror (FSM), compensating for angle-of-arrival fluctuations due to atmospheric turbulence and dynamical pointing errors in the kHz regime caused by mechanical vibrations of the telescope.

\section*{Acknowledgements}
We acknowledge funding from the Austrian  Research  Promotion  Agency (FFG) Quantenforschung und -technologie (QFTE) Contract 870003, Austrian Science and Applications Programme (ASAP) Contract 854022 and Contract 866025 and from the Austrian  Federal  Ministry  of  Education,  Science  and Research (BMBWF) and the University of Vienna via the project QUESS. We also gratefully received funding from ESA European Space Agency Contract 4000112591/14/NL/US and Contract 4000114938 /15/NL/RA/zk.

\section*{Author Contributions}
F.S., T.S., A.Z. and R.U. conceived the project. S.E. and F.S. designed the entanglement source and Alice's detection module under R.U. guidance. M.F. operated the transmitter telescope. J.H., M.F., D.R. and T.S. designed Bob's detection module and operated the receiver telescope under A.Z. guidance. B.L. wrote the post-processing software and together with S.E. and T.S. evaluated all data presented here. S.E., B.L., F.S. and T.S wrote the first draft of the manuscript. All authors discussed the results and reviewed the manuscript.

\clearpage
\onecolumngrid

\section*{Supplementary Information}

\subsection{Link performance over time (main text fig. 3)}
The parameters of the measurement illustrated in Fig.~3 of the main text are listed in Supplementary Table \ref{tab:A}, where  $N_\text{A}$ and $N_\text{B}$ are the number of measured photon detections,  $N_\text{s}^{\text{z}}$ and $N_\text{s}^{\text{x}}$ are the sifted key lengths and  $E_\mu^\text{z}$ and $E_\mu^\text{x}$ are the extracted quantum bit error rates in the z and x basis respectively.

\begin{table}[H]
	\centering
	\caption{Experimental parameters of data illustrated in Fig.~3 of the main text}
	\begin{ruledtabular}
		\begin{tabular}{cccccccc}
			\label{tab:A}
			\textrm{Time (s)}&
			\textrm{$N_\text{A}$ }&
			\textrm{$N_\text{B}$}&
			\textrm{$N_\text{s}^\text{z}$} &
			\textrm{$N_\text{s}^\text{x}$} &
			\textrm{$E_\mu^\text{z}$ (\%)} &
			\textrm{$E_\mu^\text{x}$ (\%)} &
			\textrm{Secure key rate (bps)}
			\\
			\hline
			$68$ & 891312444 & 442055 & 12100 &17960 & $6.595$ & $7.0657$ & 71.7972 \\
		\end{tabular}
	\end{ruledtabular}
\end{table}

\subsection{Secure key rates for different losses (main text fig.4)}

Here, additional experimental parameters of the measurement used in Fig.~4 of the main text are provided. They are listed in Supplementary Table \ref{tab:B}, where $\eta_\text{A}$ is the heralding efficiency and $R_\text{A}$ is the single photon rate of the source after single-mode coupling for Alice's channel, $e_\text{d}$ is the misalignment error rate, $\tau_{\text{cw}}$ is the coincidence window and $Y_\text{0A}$ and $Y_\text{0B}$ are the dark count rates per coincidence window for Alice and Bob respectively. 

\begin{table}[H]
	\caption{Experimental parameters of data illustrated in Fig.~4 of the main text}
	\begin{ruledtabular}
		\begin{tabular}{cccccc}
			\label{tab:B}
			\textrm{$\eta_\text{A}$ (\%)}&
			\textrm{$R_\text{A} (s^{-1})$ }&
			\textrm{$e_\text{d}$ (\%)}&
			\textrm{$Y_\text{0A}$}&
			\textrm{$Y_\text{0B}$} &
			\textrm{$\tau_{\text{cw}}$ (ns)}
			\\
			\colrule
			$31.42$ & $1.57 \times 10^7$ & $3.3$ & $4.71 \times 10^{-4}$ & $2.2 \times 10^{-6}$ & 1.0 \\
		\end{tabular}
	\end{ruledtabular}
\end{table}

\subsection{Secure key rate for different photon pair rates (main text fig. 5)}

In Fig. 5 of the main text we study the influence of the photon pair rate emitted by the source on the secure key rate under different channel losses. The photon pair rate per coincidence window $\mu$ is calculated via

\begin{equation}
\label{equ:Appmu}
\mu=\frac{R_\text{A}-{Y_\text{0A}}/{\tau_{\text{cw}}}}{\eta_\text{A}}\tau_{\text{cw}}
\end{equation}

from experimental data. We use 5 measurements to demonstrate the influence of the photon pair rate on the extracted secure key rate. The most relevant parameters of these measurements are listed in Supplementary Table \ref{tab:C}.

\begin{table}[H]
	\caption{Experimental parameters for different measurements illustrated in Fig.~5 of the main text}
	\centering
	\begin{ruledtabular}
		\begin{tabular}{cccccccc}
			\label{tab:C}
			
			Total Time (s) & $\mu$ & Channel Loss (dB) & $N_\text{s}^\text{z}$ & $N_\text{s}^\text{x}$ & $E_\mu^\text{z}$ & $E_\mu^\text{x}$ & Secure key rate (bps)\\ \hline
			15  & 0.0402 & 43.52  & 6062 & 8973 & 0.057737 & 0.057617 & 300.865 \\ \hline
			131 & 0.0151 & 43.75 & 19110 & 26719 & 0.027316 & 0.077772 & 114.9127 \\ \hline
			85  & 0.0493 & 46.89  & 20066 & 29351 & 0.059653 & 0.063269 & 153.526 \\ \hline
			57  & 0.0402 & 48.44 & 7846 & 11646 & 0.065893 & 0.066632 & 77.07702 \\ \hline
			100  & 0.0673 & 49.56 & 17311 & 25885 & 0.080065 & 0.080471 & 48.733 \\ 
		\end{tabular}
	\end{ruledtabular}
\end{table}

\subsection{Dual downlink from a LEO (main text fig. 6)}

We investigated the achievable secure key rate in a dual downlink scenario from a low earth orbit (LEO) in Fig.~6 of the main text. The dark count rate $Y_0$ of the detectors, which takes care of both after-pulsing from avalanche photo diodes and accidental coincidence counts is given by $Y_0 = \mu \eta P_{\text{a}} +\tau_{\text{cw}}R_{\text{DC}}$, where $\eta$ is the overall detection efficiency, $P_\text{a}$ is the after-pulsing probability and $R_\text{DC}$ is the accidental count rate of the detectors. In accordance with the measurements on our free-space link, we choose this parameters as $P_\text{a}=\unit[3.0]{\%}$, $R_\text{DC}=\unit[1000]{cps}$ and  $e_\text{d}=\unit[1.5]{\%}$.
The standard deviation of the coincidence histogram is assumed to be \unit[770]{ps}, and in accordance with the ``Micius'' dual downlink experiment \cite{Yin:2017} we choose a coincidence window of \unit[$\tau_{\text{cw}} = 2.5$]{ns}. With a heralding efficiency of \unit[30]{\%}, the entangled source on ``Micius'' produces
photon pairs at a rate of \unit[5.9]{MHz}, which corresponds to a photon pair rate per coincidence window of $\mu_{\mathrm{Micius}}=0.0492$.

\subsection{Optimal pair rate for a dual downlink scenario with SNSPDs}
In the Discussion section of the main text we mention the optimal pair rate in a dual downlink scenario assuming that the ground stations are equipped with superconducting nanowire single-photon detectors (SNSPDs). The optimal pair rate of \unit[814]{Mcps} is calculated using the link specifications listed in Supplementary Table \ref{tab:E}.

\begin{table}[H]
	\caption{Typical parameters of SNSPDs used to estimate the optimal pair rate}
	\centering
	\begin{ruledtabular}
		\begin{tabular}{ccccc}
			\label{tab:E}
			Total channel loss (dB) & Detector dead-time (ns) & $Y_\text{0A}$/$Y_\text{0B}$ & Timing jitter (ps) & $\tau_{\text{cw}}$ (ps) \\\hline 
			70 & 25 & $3.3 \times10^{-9}$ & 20 & 66.6\\
		\end{tabular}
	\end{ruledtabular}
\end{table}

\subsection{Sky noise of a typical LEO satellite pass}
We measured the background noise of the night sky with the same detection module used for our terrestrial free-space link. The module is mounted on the Ritchey-Chr\'etien reflector telescope of the optical ground station with a primary aperture of \unit[1]{m}, which followed a polar satellite orbit, depicted in Supplementary Fig.~\hyperref[fig:darks]{\ref*{fig:darks}b)}. In order to reduce the sky noise, an interference filter with a FWHM bandwidth of \unit[3]{nm} and a long pass filter were mounted in front of each detector. All detectors had a field of view of \unit[300]{$\mu$rad} at FWHM. In Supplementary Fig.~\hyperref[fig:darks]{\ref*{fig:darks}a)} the summed up sky noise of all four detector channels is illustrated, with detector dark counts subtracted. The simulated satellite pass started at an elevation angle of 8$^\circ$, went up to 40$^\circ$ at the closest approach and was terminated at the descend at 5$^\circ$.

\begin{figure}[H]
	\centering
	\includegraphics[width=\textwidth]{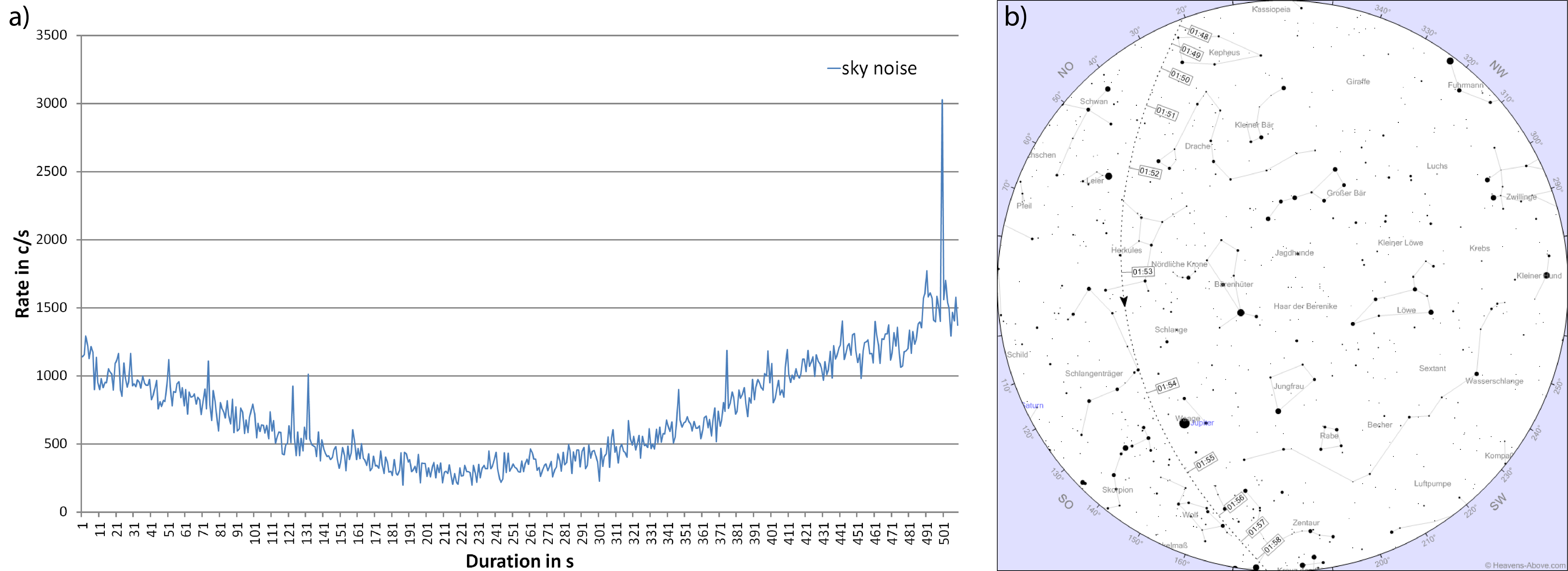}
	\caption{a.) Background sky noise of a typical LEO satellite pass summed up over all 4 detectors. The count rate at the beginning and at the end of the pass is higher due to a higher observational air mass. The spikes originate from stars crossing the field of view of the detectors. Depicted in b.) is the satellite pass (Figure taken from www.heavens-above.com).}
	\label{fig:darks}
\end{figure}

\end{document}